\documentclass[12pt,rotate]{article}
\usepackage{epsfig}

\input epsf.sty

\newcommand{\insertfig}[2]{\mbox{\epsfxsize=#1cm \epsfbox{#2.eps}}}

\newcommand{\Bx}{x_{\rm B}}

\begin{document}

\vspace{10mm}
\centerline{\large\bf MILOU : a Monte-Carlo for Deeply Virtual Compton Scattering}
\vspace{10mm}
\centerline{\bf E. Perez$^{a}$, L. Schoeffel$^{a}$, L. Favart$^{b}$}
\vspace{10mm}
\centerline{\it $^a$ CE Saclay, DAPNIA-SPP, 91191 Gif-sur-Yvette, France}
\vspace{5mm}
\centerline{\it $^b$ ULB, Boulevard du Triomphe, 1050 Bruxelles, Belgium}

\vspace{10mm}
\centerline{\bf Abstract}
In this note, we present a new generator for
Deeply Virtual Compton Scattering processes. This generator is based on 
formalism of Generalized Partons Distributions evolved at Next Leading Order (NLO).
In the following we give a brief description of this formalism and we explain the main features
of the generator for the elastic reactions, as well as for proton dissociation.
\vspace{10mm}

\section{Introduction}
\label{pheno}

\begin{figure}[h]
\begin{center}
\epsfig{figure=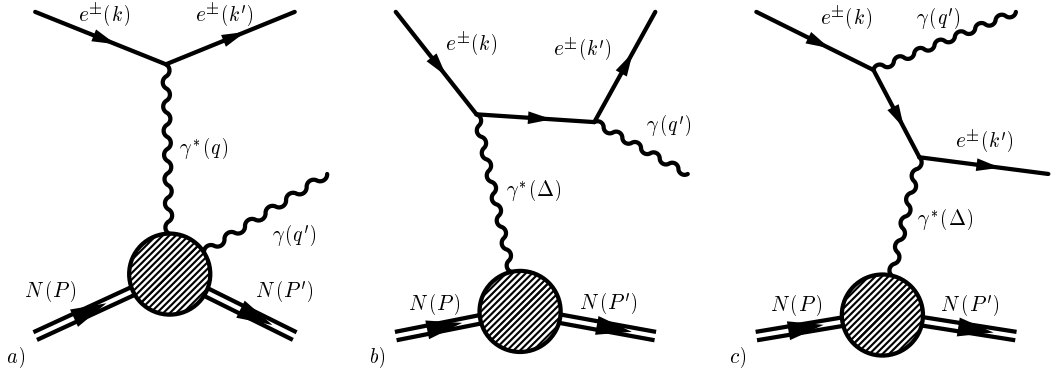,height=2.5in}
\end{center}
\caption{
{\small The contributing diagrams for the reaction $ep \rightarrow e \gamma p$ :
(a) diagram for the Deeply Virtual Compton Scattering process ; (b) and (c) diagrams for Bethe-Heitler process.}
}
\label{fig1}
\end{figure}

Deeply Virtual Compton Scattering (DVCS) is the exclusive
production 
of a real photon in diffractive $lepton-hadron$ interactions,
$l+N \rightarrow l+\gamma+ N$, as shown in Fig. \ref{fig1} (a). 
This process is calculable in perturbative QCD (pQCD),
when the virtuality, $Q^2=-q^2$, of the exchanged photon is large
and it interferes with the purely electro-magnetic Bethe-Heitler (BH) reaction
presenting the same final state (Fig. \ref{fig1} (b) and (c)). \\

The first measurements of the DVCS process at high
$W^2=(q+P)^2$~\cite{paperh1,paperzeus} and its beam-spin asymetry
in polarised $ep$ scattering at low
$W^2$~\cite{paperhermes,paperclas} have recently become available.
A proposal for dedicated studies
at the COMPASS experiment (in $\mu p$ collisions) is also under review. \\

The DVCS reaction can be regarded as the elastic scattering of the
virtual photon off the proton via a colourless exchange.  The pQCD
calculations assume that the exchange involves two partons, having
different longitudinal and transverse momenta, in a colourless
configuration. These unequal momenta are a consequence of the mass
difference between the incoming virtual photon and the outgoing real
photon. The DVCS cross section depends, therefore, on the Generalised
Parton Distributions
(GPD), which carry
information about parton correlations inside a nucleon. \\

First predictions for DVCS cross sections were based on
LO calculations \cite{ffs}. These calculations explicitely involve
the variable R defined as 
the ratio of the imaginary part for the amplitudes for 
DIS and DVCS processes : 
$$
R=\frac{Im(A(\gamma^* p \rightarrow \gamma^* p))}
{Im(A(\gamma^* p \rightarrow \gamma p))}
$$
A value of $R \simeq 0.55$ 
has been shown to be in good agreement with HERA results.
Previous Monte-Carlo (MC) for DVCS generation are based on this approach
\cite{thesisrainer,gendvcs}. We refer to this prediction as FFS
(Frankfurt-Freund-Strikman)
in the following \cite{ffs}. \\

However, even if this effective LO prediction reproduces correctly the
experimental data, it is not sufficient as it does not provide a 
direct insight of the rich information present in GPDs.
The MC described in this note has been developped to allow experimental
measurements to be compared with GPD models and to study asymetries. \\

GPDs have been studied extensively in recent years 
\cite{belitsky,freund1,diehl}.
These distributions are not only the basic, non-perturbative
ingredient in hard exclusive processes such as DVCS or exclusive
vector meson production, but they are also generalizations of the well known
Parton Distribution Functions (PDFs) from inclusive reactions. \\

In the approach developped in reference \cite{belitsky},  the scattering amplitude
is simply given by the convolution of a hard scattering coefficient
computable to all orders in perturbation theory with one type
of GPD carrying the non-perturbative information. 
The formalism for the NLO QCD evolution equations for GPDs can be found
in reference \cite{belitskyfreund1}. \\
 
Higher twists contributions to the DVCS amplitude
have been calculated in reference \cite{belitsky}.
They consist in terms
of order $O(m_N/Q)$, $O(\sqrt{-t}/Q)$ and $O(\lambda_{QCD}/Q)$ with $m_N$ the
proton mass and $t = (P-P')^2$ the momentum transfer to the outgoing
proton. They have been shown to be sizeable at low $Q^2$ within the kinematic domain
relevant for HERMES, CLAS or COMPASS experiments.

\section{Structure of the MC program}
\label{structure}

The core of the program consits of several fortran routines provided
by A. Freund. The calculations of cross sections from GPDs are done
in two steps : 
\begin{itemize}
\item[1.] The GPDs $H,{\tilde H},E,{\tilde E}$ 
are evolved at NLO by an independent code \cite{code} which provides
tables for the real and imaginary parts of the so called 
Compton Form Factor (CFFs) : at LO, they are just the convolution of GPDs
by the coefficient functions \cite{belitsky}. For example 
$$
{\cal H} (\xi,Q^2,t) = {\displaystyle \sum_{u,d,s}} 
\int_{-1}^{1} [ \frac{e_i^2}{1-x/\xi -i\epsilon}
\pm \{ \xi \rightarrow -\xi \} ] H_i(x,\xi,Q^2,t) dx
$$
with $e_i$ the fractional quark charge and $\xi = x_B / (2-x_B)$ the skewing variable.
There are different tables for each GPD and for LO/NLO approximations, as well as for
the twists-3 corrections. The GPDs parameterisations at the initial scale (in $Q^2$)
are described in reference \cite{belitsky}.

In the DGLAP domain, they consist in 
distributions based on the forward CTEQ6 parameterisations \cite{cteq}, with no 
external skeewing added in a profile function.
In each table, there are 48 bins 
in $x_B=\frac{Q^2}{2Pq}$ extending from
$10^{-4}$ till $0.7$ and for each of them 40 bins in $Q^2$ from 
$1$ GeV$^2$ till $10^4$ GeV$^2$ (regularly spaced in $\ln Q^2$). We obtain the
values for the real and imaginary parts of the CFFs considered after a 2D
spline of the $x_B-Q^2$ grid. 
\item[2.] From these CFFs, the cross sections for DVCS
and BH-DVCS interference are calculated according to formulae of reference \cite{belitsky}
(see section \ref{secxs}). The BH cross section is also calculated according to standard
expressions. During the iterative integration step, the code provides an output with the estimated
cross-section and the accuracy of the integration. It is always important to check 
the proper convergence of the integrals and that the final accuracy
is small (below $1 \%$), which means that no 
problem have occured in the integration over the kinematic range considered.
\end{itemize}

The squared amplitudes are first integrated within the kinematic domain
defined by the user.
Events are  then generated according to the differential cross sections.
These two steps make use of the BASES/SPRING package \cite{basespring2}.
For each accepted event, the kinematics of the final state particles is
stored into a PAW ntuple. The relevant kinematic formuae are detailed in
section \ref{kinematics}. 
The generated events are stored in the standard LUJETS common of the PYTHIA \cite{pythia} program.

\section{Cross section formulae and kinematics}
\label{secxs}

In the MC, we are calculating the five-fold cross section for the
DVCS process 
\begin{eqnarray}
\label{WQ}
\frac{d\sigma}{d\Bx dy d|t| d\phi d\varphi}
=
\frac{\alpha^3  \Bx y } {16 \, \pi^2 \,  {Q}^2 \sqrt{1 + \epsilon^2}}
\left| \frac{\cal T}{e^3} \right|^2 
\end{eqnarray}
in which the amplitude ${\cal T}$ is the sum of the DVCS and
Bethe-Heitler (BH) amplitudes : 
${\cal T}={\cal T}_{\rm DVCS}+{\cal T}_{\rm BH}$.
This cross section depends on the Bjorken variable $\Bx$, the squared
momentum transfer $t = (P - P')^2$, the lepton energy fraction $y
= P\cdot q/P\cdot k$, and, in general, two
azimuthal angles. We use the convention
$\epsilon \equiv 2 \Bx \frac{m_N}{{Q}}$. \\

\begin{figure}[t]
\begin{center}
\mbox{
\begin{picture}(0,250)(250,0)
\put(70,0){\insertfig{13}{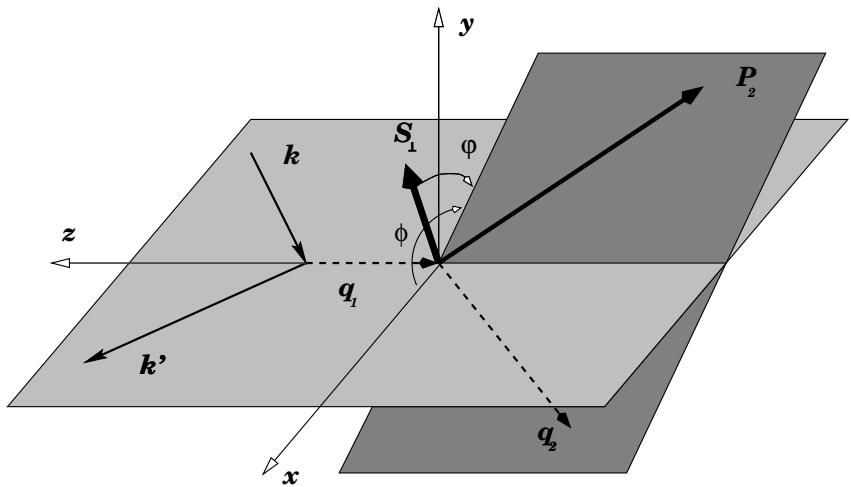}}
\end{picture}
}
\end{center}
\caption{\label{Fig-Kin}
{\small
The kinematics of the leptoproduction in the target rest frame. 
In this figure taken from reference \cite{belitsky}, $k$, $k'$, $P_2$ and $q_2$ are respectively the four-momenta of
the initial lepton, the scattered lepton, the outgoing proton and the real photon.
The $z$-direction is chosen counter-along the three-momentum of the incoming
virtual photon. The lepton three-momenta form the lepton scattering plane,
while the recoiled proton and outgoing real photon define the hadron
scattering plane. In this reference system the azimuthal angle of the
scattered lepton is $\phi_l = 0$, while the azimuthal angle between the
lepton plane and the recoiled proton momentum is $\phi_N = \phi$. When the
hadron is transversely polarised (in this reference frame) $S_\perp = (0,
\cos {\mit\Phi}, \sin {\mit\Phi}, 0)$, the angle between the polarisation
vector and the scattered hadron is denoted as $\varphi = {\mit\Phi} -
\phi_N$. }}
\end{figure}

In equation (\ref{WQ}), $\phi = \phi_N - \phi_l$ is the angle between the lepton
and hadron scattering planes and $\varphi = {\mit\Phi} - \phi_N$ is the
difference of the azimuthal angle $\mit\Phi$ of the transverse part of the
nucleon polarisation vector $S$, i.e., $S_\perp = (0, \cos{\mit\Phi},
\sin{\mit\Phi}, 0)$, and the azimuthal angle $\phi_N$ of the recoiled
hadron. Our frame is rotated with respect to the laboratory one in such a
way that the virtual photon four-momentum has no transverse components, see
Fig.\ \ref{Fig-Kin}. For the kinematics we choose the following convention :
the $z$-component of
the virtual photon momentum is negative and $x$-component of
the incoming lepton is positive with $k = (E, E \sin\theta_l, 0, E \cos\theta_l )$, $q =
(q_0, 0, 0,-|q_3|)$. Other vectors are $P = (M, 0, 0, 0)$ and $P' =
(E_2, |\mbox{\boldmath$P$}_2| \cos\phi \sin\theta_N, |
\mbox{\boldmath$P$}_2| \sin\phi \sin\theta_N, |\mbox{\boldmath$P$}_2|
\cos\theta_N)$. 

\subsection{Cross sections}

The BH amplitude is real (to
the lowest order in the QED fine structure constant) and is parameterised in
terms of electromagnetic form factors, which we assume to be known from other
measurements. \\

According to reference \cite{belitsky},
the BH, DVCS and interference terms of equation (\ref{WQ}), namely  $|{\cal T}_{\rm BH}|^2$, 
$|{\cal T}_{\rm DVCS}|^2$ and  ${\cal I}
={\cal T}_{\rm DVCS} {\cal T}_{\rm BH}^* + {\cal T}_{\rm DVCS}^* {\cal T}_{\rm BH}$
can be written :
 
\begin{eqnarray}
\label{Par-BH}
&&|{\cal T}_{\rm BH}|^2
= \frac{e^6}
{\Bx^2 y^2 (1 + \epsilon^2)^2 \Delta^2\, {\cal P}_1 (\phi) {\cal P}_2 (\phi)}
\left\{
c^{\rm BH}_0
+  \sum_{n = 1}^2
c^{\rm BH}_n \, \cos{(n\phi)} + s^{\rm BH}_1 \, \sin{(\phi)}
\right\} 
\\
\label{AmplitudesSquared}
&& |{\cal T}_{\rm DVCS}|^2
=
\frac{e^6}{y^2 {\cal Q}^2}\left\{
c^{\rm DVCS}_0
+ \sum_{n=1}^2
\left[
c^{\rm DVCS}_n \cos (n\phi) + s^{\rm DVCS}_n \sin (n \phi)
\right]
\right\} 
\\
\label{InterferenceTerm}
&&{\cal I}
= \frac{\pm e^6}{\Bx y^3 \Delta^2 {\cal P}_1 (\phi) {\cal P}_2 (\phi)}
\left\{
c_0^{\cal I}
+ \sum_{n = 1}^3
\left[
c_n^{\cal I} \cos(n \phi) +  s_n^{\cal I} \sin(n \phi)
\right]
\right\} 
\end{eqnarray}
where the $+$ ($-$) sign in the interference stands for the negatively
(positively) charged lepton beam.
In this expression, the Fourier coefficients $c_i$, $s_i$ are fonction of CFFs
(and then GPDs) and $1/{\cal P}_1$ and $1/{\cal P}_2$ are the lepton
BH propagators. \\

We have mentioned above that the real and
imaginary parts of the CFFs (convolution of the GPDs by a hard coefficient function)
are given in tables of $x_B-Q^2$, thus the $t$ dependence is assumed to be
factorised and has to be defined by the MC user. We have included an option
in the steering which allows to consider a global exponential behaviour of the
cross section 
$$
d\sigma/dt \propto exp(B(Q^2) \ t)
$$
in which the $t$ slope $B(Q^2)$
can be chosen as a linear function of $\ln Q^2$.
Alternatively, the $t$ dependence can also be given by the
Pauli-Dirac form factors for each GPD flavor.

\subsection{Event kinematics}
\label{kinematics}

The reconstruction of energies and angles of the scattered positron, 
photon and proton is carried out for each event from $x$, $Q^2$, $t$
and $\phi$. \\

The kinematics of the outgoing lepton (and then of the
virtual photon) is trivially obtained in the laboratory frame.
The kinematics of the outgoing nucleon 
$$
P' =
(E_2, |\mbox{\boldmath$P$}_2| \cos\phi \sin\theta_N, |
\mbox{\boldmath$P$}_2| \sin\phi \sin\theta_N, |\mbox{\boldmath$P$}_2|
\cos\theta_N)
$$
is most easilly obtained in the
the frame of Fig. \ref{fig1}, where the target nucleon is at rest.
We use 
$$
E_2 = m_N - \frac{t}{2m_N}
\ \ \ \ \
|\mbox{\boldmath$P$}_2| = \frac{\sqrt{t(t-4m_N^2)}}{2 m_N}
$$
and
$$
\cos \theta_N = \frac{(q_3)^2-(q_0)^2+|\mbox{\boldmath$P$}_2|^2}
{2 |\mbox{\boldmath$P$}_2| \  q_3}
$$
in which the virtual photon components can be expressed as
$$
q_0 = \frac{t+Q^2/x_B}{2 m_N} 
\ \ \ \ \
q_3 = -\frac{Q^2}{2 m_N x_{B}} \sqrt{1+\frac{4 m_N^2 x_{B}^2}{Q^2}}
$$

After a Lorentz transformation, we finally get all variables 
(energies and angles of incoming and outgoing particles)
in the
laboratory frame. In the ntuple produced during the generation,
these variables are given in both frames.

%
%

%
%

\section{Integration and Generation}
\label{basespring}
Integration and generation are realized with the BASES/SPRING package \cite{basespring2}.
By BASES, probability distributions are calculated by integrating the differential cross-section
over the phase space and saved in a file which is then used in the generation step.
By SPRING, events are generated by the MC method according to the above distributions.

\section{Proton dissociation}
\label{spdiss}

It has been shown that the proton dissociation is a non negligible
contribution (10\% to 20\%) in the measured DVCS samples for H1 and ZEUS experiments
\cite{paperh1,paperzeus}. Thus, it is important to include this process
in the MC. \\

In case of proton dissociation the DVCS process $ep \rightarrow e Y \gamma$
leads to a state of mass $M_Y$ and the generation of the $M_Y$ spectrum follows
the parameterisation exposed in \cite{goullianos}, as it is done in the DIFFVM
MC \cite{diffvm}. The main hypothesis is that  $\frac{d\sigma}{dM_Y^2}$
can be factorised with the elastic DVCS cross-section, with :
$$
\frac{d\sigma}{dM_Y^2} = \frac{f(M_Y^2)}{M^{2(1+\epsilon)}}
$$
In the continuum region ($M_Y^2 > 3.6$ GeV$^2$), $f(M_Y^2)=1$ leading to a 
global $M_Y$ dependence in $1/M^{2(1+\epsilon)}$ for proton dissociation.
In the resonance region ($M_Y^2  < 3.6$ GeV$^2$), $f(M_Y^2)$ is the result
of a fit for the proton diffractive dissociation on deuterium (at fixed $t$)
\cite{goullianos}. It is important to note that this treatment is only acceptable
for DVCS at large $W$ ($W > 20$ GeV). Thus, it must not be applied in the BH mode
and in the low $W$ domain, where the stucture of resonances is more complex.

\section{Examples}

\subsection{Elastic DVCS process}

\begin{figure}[h]
\begin{center}
\epsfig{figure=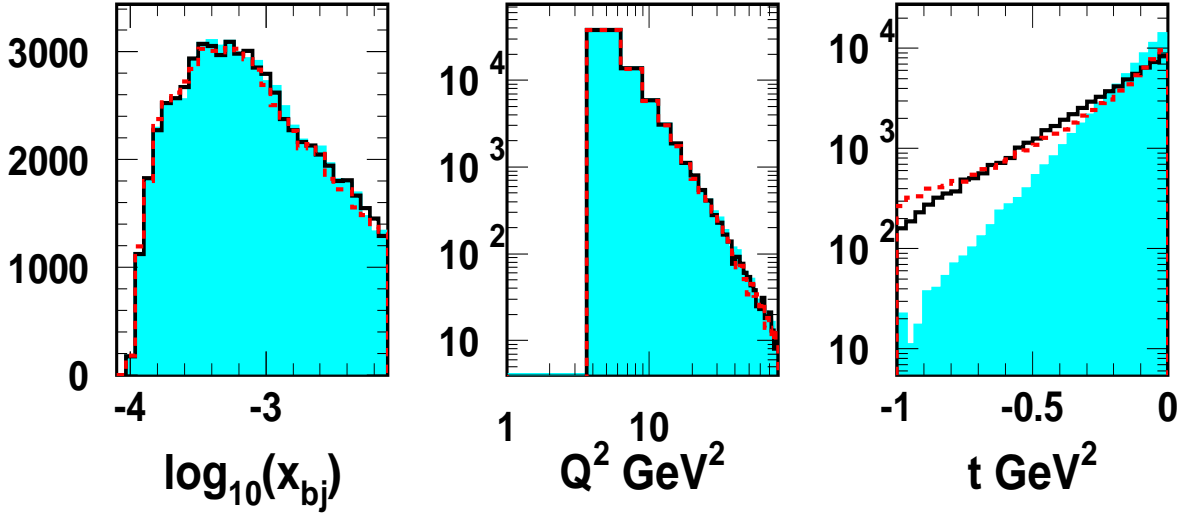,height=10cm,width=18cm}
\end{center}
\caption{
{\small
Predictions for the kinematic variables at NLO of the DVCS MC for different behaviours in $t$ :
the full histogram stands for a global exponential $t$ dependence
of the DVCS cross section 
in $e^{B \ t}$ with $B=7$ GeV$^{-2}$ ; the full line represents 
also an exponential $t$ dependence with a slope of $B=4$ GeV$^{-2}$
and the dotted line has been obtained when considering the
Dirac-Pauli form factors for each GPDs. Histograms are normalized to the number of
events.
}}
\label{f1_examples}
\end{figure}

\begin{figure}[h]
\begin{center}
\epsfig{figure=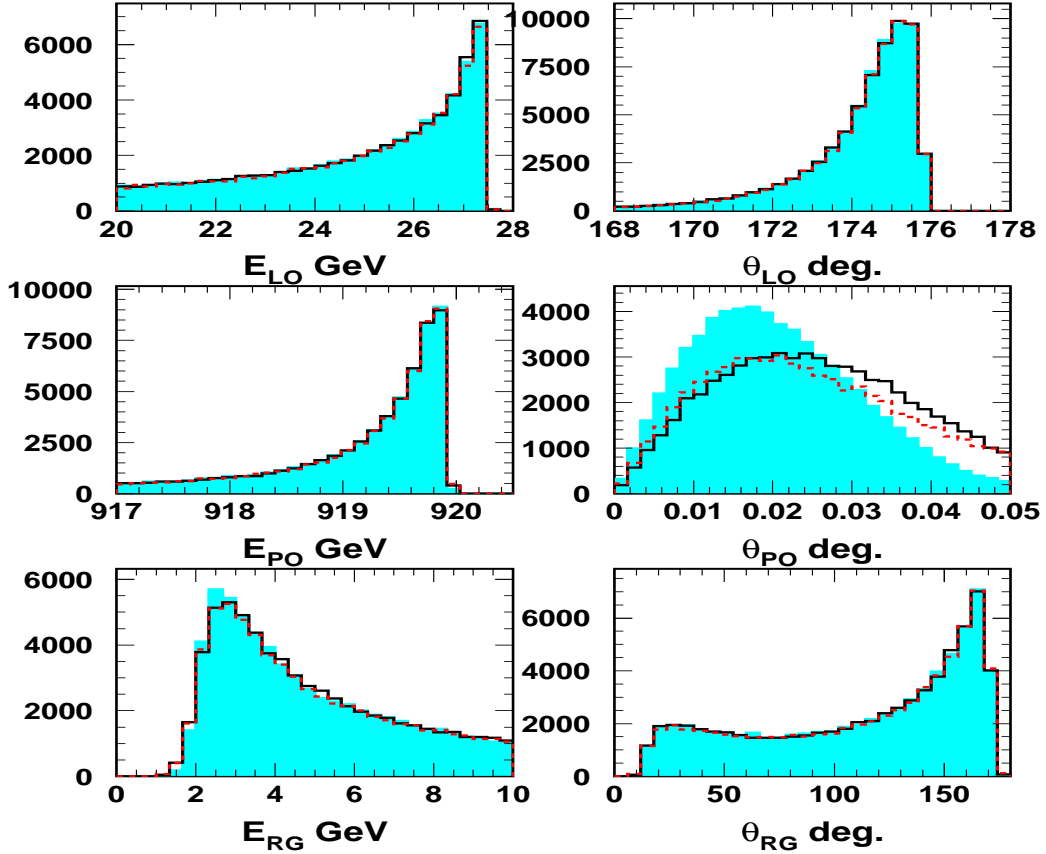,height=12cm,width=15cm}
\end{center}
\caption{
{\small
With the same convention as in Fig. \ref{f1_examples}, (different
t dependences), we represent the energies and angles of the 
scattered (outgoing) lepton $E_{LO}$ and $\theta_{LO}$,
for the outgoing proton $E_{PO}$ and $\theta_{PO}$ and for 
the real photon $E_{RG}$ and $\theta_{RG}$.
}}
\label{f2_examples}
\end{figure}

To illustrate the ouput of the DVCS MC, we first present some examples in
the elastic case ($e p \rightarrow e p \gamma$). In Fig. \ref{f1_examples}
we present the generated kinematic variables in the range 
$10^{-4} \le x_{Bj} \le 0.1$, $4 \le Q^2 \le 100$ GeV$^2$ and
$t>-1$ GeV$^2$. The generation is done at NLO for different $t$
dependences of the GPDs as explained in the caption of Fig. \ref{f1_examples}.
For the same generation, Fig. \ref{f2_examples} represents the spectra for 
energies and polar angles of the outgoing particles.
The normalization is done to the number of events.

\begin{figure}[h]
\begin{center}
\epsfig{figure=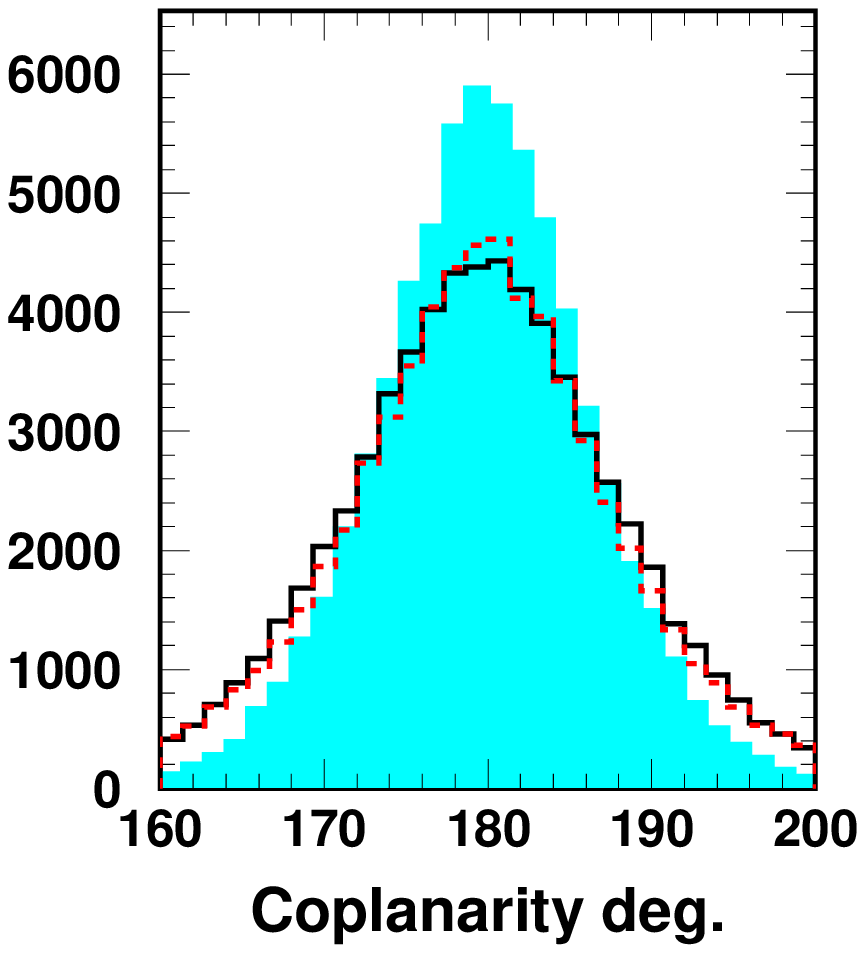,height=4.cm,width=7cm}
\end{center}
\caption{{\small Coplanarity : absolute difference of the azimuthal angles of the
scattered lepton and real photon, for the three cases mentioned in Fig. \ref{f1_examples}.}}
\label{f3_examples}
\end{figure}

In Fig. \ref{f3_examples}, we show the coplanarity 
(absolute difference of the azimuthal angles of the
scattered lepton and real photon) of the elastic DVCS process
(again with the three different $t$ behaviours considered since Fig. \ref{f1_examples}).
We observe the broadening of this distribution when the $t$ slope becomes larger.
Indeed, for small momentum transfer the $e \gamma$ system is balanced and the 
coplanarity is close to $180^o$, whereas for larger momentum transfer the
$e \gamma$ system becomes unbalanced and a deviation form $180^o$ is observed.
And, of course, the fraction of events with a large momentum transfer is
enhanced when $B$ becomes smaller.

\begin{figure}[!]
\begin{center}
\epsfig{figure=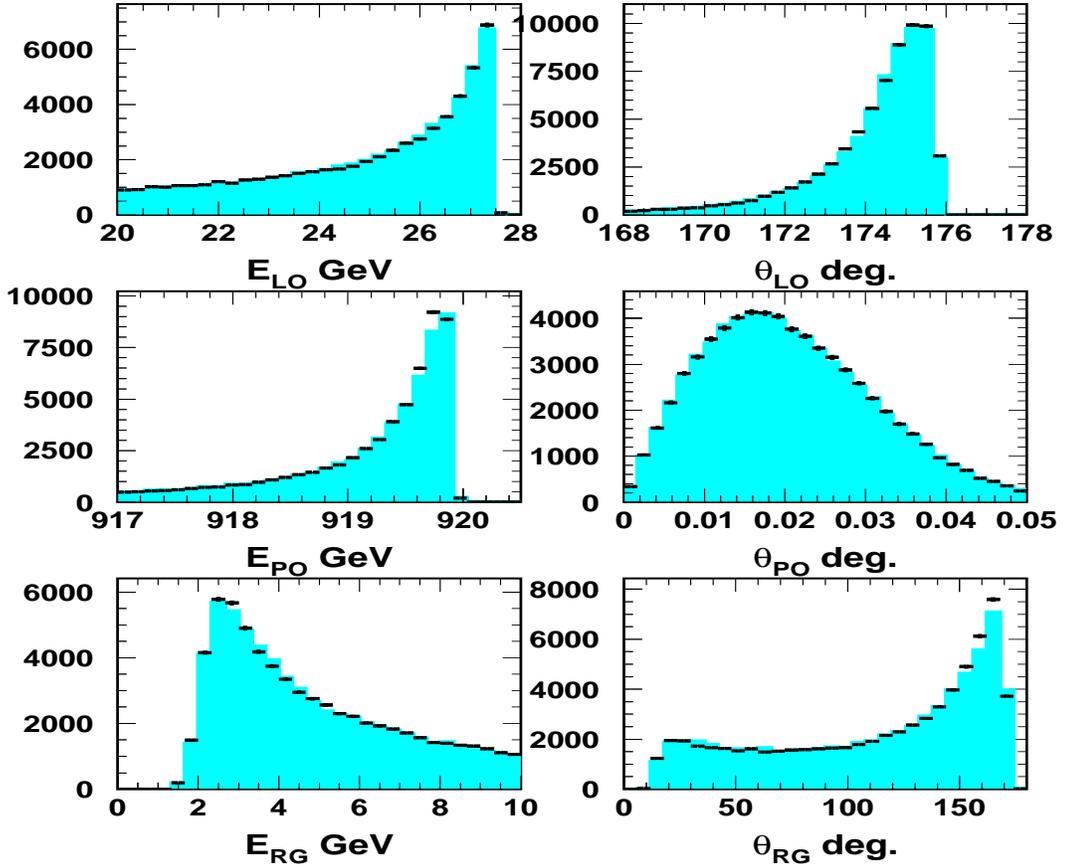,height=12.cm,width=15cm}
\end{center}
\caption{{\small Comparisons of the predictions based on the LO calculations
\cite{ffs} with a value of $R=0.55$ (black points) and
of the NLO calculations (full histogram). In both cases, we have 
considered a global exponential $t$ dependence with $B=7$ GeV$^{-2}$.
Histograms are normalized to the number of
events.
}}
\label{f4_examples}
\end{figure}

As mentioned in section \ref{pheno}, previous measurements (and then acceptance
corrections) at low $x_B$ \cite{paperh1,paperzeus}
are based on the FFS approximation for the DVCS cross-section \cite{ffs}.
On Fig. \ref{f4_examples}, we 
compare the FFS (for R=0.55) and NLO predictions of the DVCS MC in the same generated kinematic
range : $10^{-4} \le x_{Bj} \le 0.1$, $4 \le Q^2 \le 100$ GeV$^2$ and
$t>-1$ GeV$^2$. For this comparison we only consider
a global exponential $t$ 
dependence with $B=7$ GeV$^{-2}$. Histograms are normalized
to the number of events and we notice a good agreement in shapes. However
the cross section calculated in the NLO case is calculated to be about $15 \%$ higher
than the LO FFS one (for the GPDs parameterisations included in the code).
Regarding the good agreement in shapes, one can wonder whether it is useful to
extend the MC from FFS LO to NLO GPDs calculations. However, it is important to
note at this stage that GPDs at NLO become essential to predict asymetries as explained
later in this note. Also, as mentioned previously, with GPDs we can consider
different $t$ dependences for the different flavors, which is obviously not possible
in the FFS approximation.
 
\vfill 
\clearpage

\subsection{DVCS vs BH cross-sections}

\begin{figure}[h]
\begin{center}
\epsfig{figure=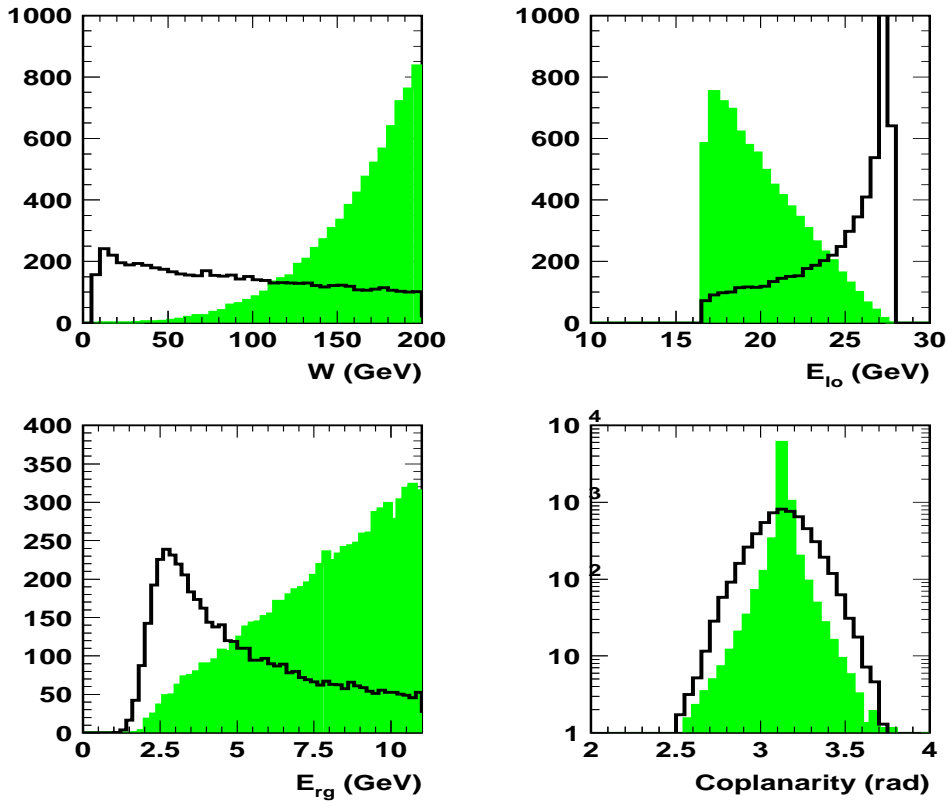,height=12cm,width=15cm}
\end{center}
\caption{{\small Spectra of DVCS (full line) compared to BH (full histogram) 
processes for $W$, energies of the outgoing lepton and
the real photon and the coplanarity. 
The configuration considered for these plots is the one with the lepton scattered backward
($\theta_{lo} > 160^o$) and  the real photon scattered with  $\theta_{rg} < 160^o$.
Histograms are normalized to luminosities
as calculated by the MC. 
}}
\label{fnew1}
\end{figure}

In Fig. \ref{fnew1} we present another illustration of the MC : we compare the DVCS with
the BH process when the lepton is scattered backward ($\theta_{lo} > 160^o$),
the real photon is scattered with  $\theta_{rg} < 160^o$ and the azimuthal angle is integrated over.
This configuration
is the one analysed in H1 and ZEUS experiments to extract the DVCS cross-sections
\cite{paperh1,paperzeus}. Histograms (Fig. \ref{fnew1}) are normalized to 
luminosities and the interference between DVCS and BH is calculated to be
negligible : as the azimuthal angle is integrated over, the constant term in the expression
for the interference (see formula \ref{InterferenceTerm}) is the only (negligible) contribution.
We notice the different behaviours in $W$, $E_{lo}$,
which allow a good separation power at low $W$ ($W \le 120$ GeV) and large $E_{lo}$
($E_{lo} \ge 15$ GeV)
and thus allow the measurement of the DVCS cross-section.

\begin{figure}[h]
\begin{center}
\epsfig{figure=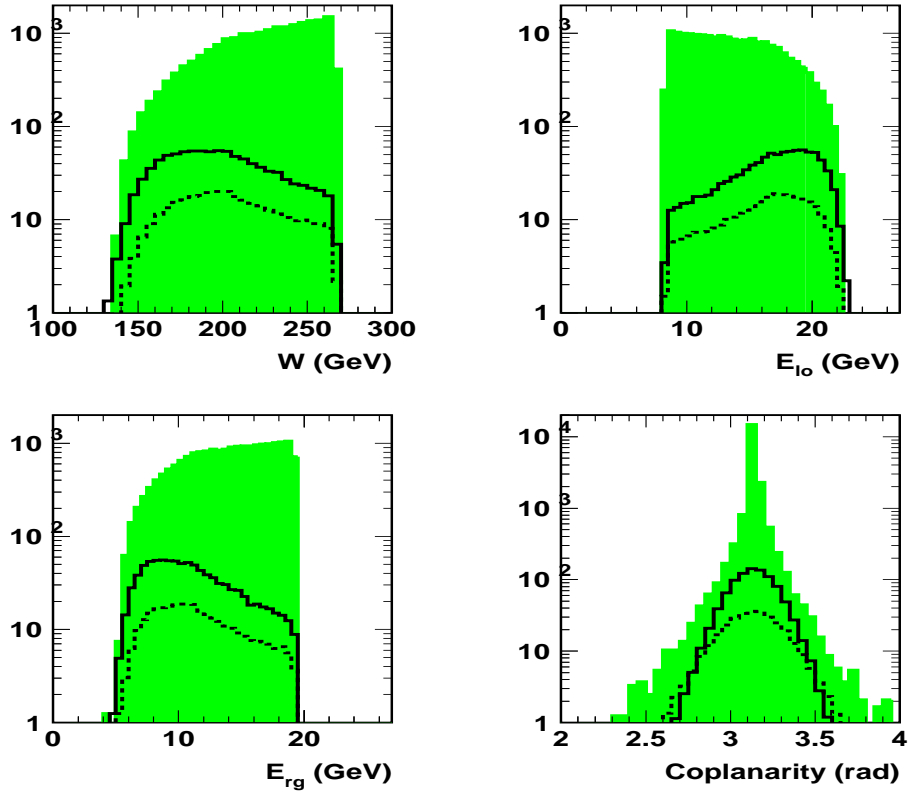,height=12cm,width=15cm}
\end{center}
\caption{{\small Spectra for DVCS (full line) compared to BH (full histogram) 
processes for $W$, energies of the outgoing lepton and
the real photon and the coplanarity. 
The configuration considered for these plots is the one with the lepton and the
real photon scattered backward
($\theta_{lo,rg} > 160^o$). 
Histograms are normalized to luminosities
as calculated by the MC. In this configuration, it is interesting to notice that
the interference (dotted line) is small but non negligible (about $0.5 \%$) of
the BH sample.  
}}
\label{fnew2}
\end{figure}

In Fig. \ref{fnew2} we compare DVCS and BH when both lepton and photon are
scattered backward ($\theta_{lo,rg} > 160^o$). Histograms are normalized to luminosities.
We notice that in this case the BH process is dominating, the DVCS representing
about $6 \%$ of the sample and the interference about $0.5 \%$.

\vfill
\clearpage

\subsection{Proton dissociation DVCS process}

\begin{figure}[h]
\begin{center}
\epsfig{figure=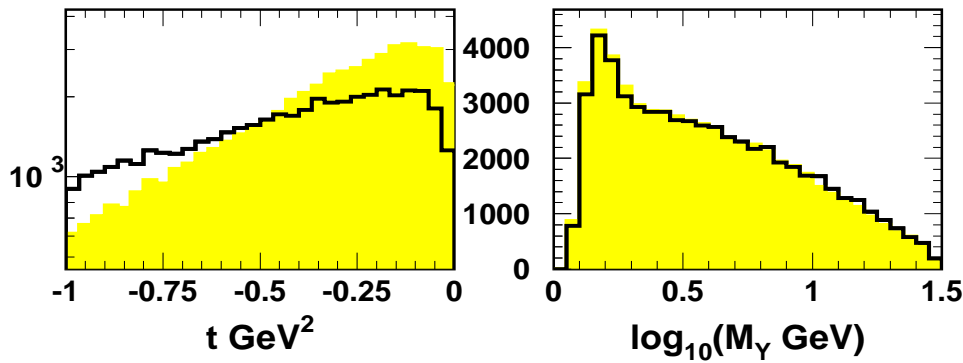,height=7cm,width=15cm}
\end{center}
\vspace*{-1cm}
\caption{{\small
Example of predictions of the DVCS MC for proton dissociation.
With two values of $B$ (slope of the global exponential $t$ dependence) :
$B=1$ GeV$^{-2}$ (black line) and $B=2$ GeV$^{-2}$ (full histogram),
we represent the $t$ spectrum and the generated mass of the
proton dissociated system $M_Y$ (in log).
}}
\label{f5_examples}
\end{figure}

As mentioned in section \ref{spdiss},
in case of proton dissociation the DVCS process $ep \rightarrow e Y \gamma$
leads to a state of mass $M_Y$. 
We present on Fig. \ref{f5_examples} the generated $M_Y$ for two values
of the $t$ slope (slope of the global exponential $t$ dependence).
As mentioned in section \ref{spdiss}, we notice the $M_Y$ behaviour in 
$$
\frac{d\sigma}{dM_Y^2} = \frac{f(M_Y^2)}{M^{2(1+\epsilon)}}
$$
with the resonance region at low $M_Y$ ($M_Y^2  < 3.6$ GeV$^2$) and the
$1/M^{2(1+\epsilon)}$ shape at larger values.

Away from the resonance region, the system $Y$ is treated as a quark-diquark (q-qq) system.
Its hadronisation is performed by PYTHIA. We have added an option
compared to the treatment done in the DIFFVM MC. In DIFFVM,
it is assumed that the proton splits into a q-qq system, so that the quark 
couples to the pomeron leaving a diquark remnant. Due the possible spin
states of this diquark system (in a singlet or triplet wave function), different
probabilities are assigned to the different configurations of the q-qq system. 
However, for a DVCS process, the coupling of the valence quark can also be done to a quark : see
for example the LO handbag diagram. Thus, the probabilities for the q-qq system describing the
proton have to be modified by taking into the account the electric charge coupling. 
We have introduced a real parameter in the steering which allows to define the type of probabilities
considered : $\alpha = 0$ if we only consider the probabilities defined by spin states $ P_0$ , $\alpha = 1$ in case
of spin states and electric coupling ${ P_1}$, and $0 < \alpha < 1$ for 
${ P_\alpha = (1-\alpha) P_0 + \alpha P_1 }$.


\section{Steering cards}
\label{steering}

\begin{description}

\item EXP    1 : 1/0 to use de logarithmic/linear binning in the kinematic range for $Q^2$ and $x_{B}$
%
%
\item SEED  2345671 : generation seed value
\item FIXED  FALSE : FALSE/TRUE to select collider/fixed-target modes
\item ELEP   -27.55 : beam momentum (in GeV/c) of the lepton beam 
\item ETARG   920. : in collider mode, beam momentum (in GeV/c) of the proton beam or, in fixed
               target mode, mass of the target 
\item LCHAR  +1. : lepton charge in units of $e$ in case TINTIN is set to FALSE
   (the code is using the grids of 
 real and imaginary parts for CFFs)
\item LPOL  0. : polarisation of lepton, +1 in direction of movement, -1 against direction of
movement, 0 for unpolarised
\item TPOL  0. :  target polarisation, +1(-1) = polarised along (opposite) probe/target beam
direction, 0 for unpolarised
\item ZTAR  1. : charge of the target 
\item ATAR  1. : atomic number of the target
\item SPIN  1. : spin of target nucleon type (in units of ${\hbar}/2$)

\item IRAD  0 : 1 for QED radiative corrections to be applied, 0 otherwise (only the ISR are included
                in the code) 
\item IELAS  1 : 1 to run the code for the elastic case only, 0 to run in the
           proton dissociation mode only. Note that we can not run both
	   cases together and, as mentioned above, the proton dissociation treatment implemented in the code
           can only be used for a pure DVCS process 
\item IRFRA  1 : in case of proton dissociation for the resonances domain, 
           IRFRA=0 if the user wants to use DIFFVM's routines
           for the decays of resonances, 1 if this user wants to decay via PYTHIA 
\item PROSPLIT  1. : treatment of dissociated proton in the continuum domain,
               PROSPLIT $\equiv \alpha$ (see above) steers how the proton is splitted into quark-diquark
\item EPSM   0.08 : for proton dissocation, EPSM $\equiv \epsilon$ with 
$d\sigma/dM_Y^2 \propto 1/M_Y^{2(1+\epsilon)}$ 
\item TINTIN  FALSE : TRUE to run in the LO FFS \cite{ffs} approximation, FALSE
                      to use the grids for CFFs
\item BTIN    2. : slope of the exponential t dependence (in the FFS approximation)
\item RTIN    0.55 : value for the variable $R=\frac{Im(A(\gamma^* p \rightarrow \gamma^* p))}
               {Im(A(\gamma^* p \rightarrow \gamma p))}$
                  (see section \ref{pheno}) 
\item F2QCD   TRUE : in the FFS approximation $\sigma_{DVCS} \propto F_2^2 / R^2$,
               this parameter determines the parameterisation considered for $F_2$ :
	       TRUE/FALSE respectively for the H1 QCD fit or ALLM results 
\item DIPOLE  FALSE : if the parameter TINTIN is  TRUE, it's possible to run the code
                using the diople model formula (in this case DIPOLE must be set to TRUE) 
\item IGEN  0 : generation mode with BASES/SPRING : 0 for the grid calculation 
 and the generation and 4 for the generation only (which requires the file 
 bases.data)
\item NGEN    10 : dummy
\item NPRINT  10 : number of events for which the PHYTHIA output is printed
\item NCALL  10000 : parameter for BASES, number of sampling points per iteration
\item ITMX1  10 : parameter for BASES, number of iterations for the grid defining step
\item ITMX2  10 : parameter for BASES, number of iterations for the integration step
\item IDEBUG  0 : debug flag
\item NXGRID 58 : Number of $X$ points in the CFFs grids
\item NQGRID 40 : Number of $Q^2$ points in the CFFs grids
\item IPRO 2 : process to generate : 1,2,3,4 or 5 for BH, DVCS, interference BH-DVCS, charge 
               asymetry or single spin asymetry respectively. 
\item IORD 2 : order for the calculations of the GPDs amplitudes : 1/2 for LO/NLO
\item XMIN    1.0e-4 : 
 lower bound of $x_{B}$ for the kinematic domain of the generation. If the MC user run the
 code with the use of CFFs grids, this value must be larger than the lower bound of the
 grids ($10^{-4}$ in our case)
\item XMAX    1.0e-1 : upper bound of $x_{B}$
\item QMIN    3.0 : lower bound for $Q^2$ (in GeV$^2$) 
\item QMAX  300.0 : upper bound for $Q^2$ (in GeV$^2$) 
\item TMIN    0.0 : upper bound for $t=(p'-p)^2$  (in GeV$^2$)
\item TMAX   -1.5 : lower bound  for $t=(p'-p)^2$  (in GeV$^2$)
%
%
%
\item MYMIN  1.13 : lower bound on $M_Y$ in proton dissociation case
\item MYMAX  30. : upper bound on $M_Y$ 
%
%
%
\item NSET   2 : integration over PHI $\equiv \phi$ : 2 for integration, 1 otherwise
\item PHI    0. : in case of NSET=1, 
azimuthal angle between lepton and final state scattering planes 
\item PPHI   0. : in case of NSET=1,
 angle between lepton plane and transverse polarisation vector
\item THETA  0. : 
angle between longitudinal and transverse polarisation
%
%
%
%
\item TWIST3  FALSE : Twists 3 calculations (TRUE) or not (FALSE)
%
%
%
\item ITFORM   0 : 
in case of the use of CFFs grids,
if ITFORM=0, the $t$ dependence is in $\exp(bt)$ with
                   $b =$ BQCST $+$ BQSLOPE $\log(Q^2/Q0SQ)$ ; if ITFORM=2, the 
                   Pauli-Dirac form factors are used for the $t$ dependence ;
                   finally,
                   if ITFORM=4,  the $t$ dependence is in $\exp(bt)$ for $x_B<$ X0DEF and Pauli-Dirac
                   form factors are considered otherwise. Note that in case
of the LO FFS approximation the $t$ dependence is set by the parameter BTIN (see above).

\item BQCST    7.0 : see ITFORM
\item BQSLOPE  0.0 : see ITFORM 
\item BGCST    7.0 : see ITFORM
\item BGSLOPE  0.0 : see ITFORM
\item Q0SQ     2.0 : see ITFORM
\item X0DEF    0.1e0 : see ITFORM
%
%
%
%
\item THLMIN   00.0 : lower bound on $\theta_{lepton}$ (in deg.)
\item THLMAX  180.0 : upper bound on $\theta_{lepton}$ (in deg.) 
\item ELMIN    10.0 : lower value of $E_{vis}$ (in GeV)
\item THGMIN   00.0 : lower bound on $\theta_{\gamma}$ (in deg.)
\item THGMAX  180.0 : upper bound on $\theta_{\gamma}$ (in deg.)
\item EIMIN     0.00001 : minimum value for the ISR photon (in GeV) 
\item YMAX   0.4 : maximum value for $y=Q^2/(sx)$
\item M12MIN 1. : minimum value for the reconstructed mass of the outgoing lepton
 and the real photon
\item ELMB   0. : minimum value of the outgoing lepton energy (in GeV)
\item EGMB   0. : minimum value of the real photon energy (in GeV)
%
%

\end{description}

\section{Summary}

MILOU is a new generator for DVCS based on
the formalism of Generalized Partons distributions.
This MC has been developped to allow experimental
measurements to be compared with GPD models (at LO or NLO) and to study asymetries. 
The generation of
BH processes and interference between DVCS and BH are also aviable. In addition,
in case of pure DVCS we have included the possiblility to study
proton dissociation.

\section{Acknowledgements}

We are very thankfull to A. Freund and M. McDermott for providing us tables
with Compton Form Factors and the core of code for DVCS cross-section calculations.
We also wish to thank A. Bruni for helpful discussion on proton dissociation.




\end{document}